\documentclass{PoS}

\usepackage{bm}

\newcommand\onelineequation[2]{%
\begin{equation}\label{#1}
#2
\end{equation}%
}
\newcommand\twolineequation[4]{%
\begin{eqnarray}
#2
\label{#1}\\
\nonumber #4
\label{#3}
\end{eqnarray}%
}
\newcommand\onefigure[4]{%
\begin{figure}[#4]
\centering
\includegraphics[width=#2]{#1}
\caption{#3}\label{#1}
\end{figure}%
}
\newcommand\twofigures[5]{%
\begin{figure}[#5]
\centering
\begin{tabular}{c c}
\includegraphics[width=#3]{#1}&\includegraphics[width=#3]{#2}
\end{tabular}
\caption{#4}\label{#1_#2}
\end{figure}%
}
\newcommand{\sectionquote}[1]{%
\begin{flushright}
\begin{tabular}{|p{0.7\textwidth}}
\textit{#1}
\end{tabular}
\end{flushright}
}
\newcommand\oh{{\textstyle\frac{1}{2}}} 
\newcommand\of{{\textstyle\frac{1}{4}}} 
\newcommand\cN{{\cal{N}}}
\title{Numerical Study of the SU(2) Yang--Mills Vacuum State\\{\Large Much Ado About Nothing?}%
}

\ShortTitle{Numerical study of the SU(2) Yang-Mills vacuum state} 

\author{Jeff Greensite\\
Physics and Astronomy Dept., San Francisco State University, San Francisco, CA 94132, USA\\
E-mail: \email{jgreensite@gmail.com}}

\author{
				\speaker{\v{S}tefan Olejn{\'\i}k}\\
        Institute of Physics, Slovak Academy of Sciences, SK--845 11 Bratislava, Slovakia\\
        E-mail: \email{stefan.olejnik@savba.sk}} 

\abstract{Numerical results for relative weights of test gauge-field configurations in the vacuum of the SU(2) lattice gauge theory in $(3+1)$ dimensions are compared with expectations following from various proposals for the Yang--Mills vacuum wave functional that interpolate between the free-field limit and the dimensional-reduction form.}

\FullConference{From quarks and gluons to hadronic matter: A bridge too far?\\
		 September 2--6, 2013\\
		 European Centre for Theoretical Studies in Nuclear Physics and Related Areas (ECT*), Villazzano, Trento (Italy)}

\begin{document}

\section{The Taming of the Shrew}\label{section1}
\sectionquote{wherein the problem to be solved is introduced that looks
simple but defies solution for years.}

	In the theory of strong interactions, quantum chromodynamics, one can dream of finding the wave functional describing its ground state (vacuum) in the Schr\"odinger representation:
\twolineequation{VWF}
{\fbox{$\Psi_0\left[u^i_A(x),d^i_A(x),s^i_A(x),c^i_A(x),b^i_A(x),t^i_A(x);A^a_\mu(x)\right]$}\quad\phantom{A}}
{}
{A=1,2,3,4;\quad i=1,2,3;\quad a=1,2,\dots,8; \quad\mu=0,1,2,3;}
that should encompass colour confinement, chiral symmetry breaking, and other observed phenomena. Even if one forgets about the subtleties of how to make such an object a mathematically well-defined entity, the problem still looks very difficult, if not utterly hopeless: in our world with six flavours of quarks with three colours, each represented by a Dirac spinor of four components, and with eight
four-vector gluons, the vacuum wave functional depends on \textit{104 fields at each point of space} (not taking gauge invariance into account).

	Still, one can simplify QCD considerably in many ways, hoping that the amputee will share (some of) the most important  features with the full theory \cite{Feynman:1981ss}. One can omit quarks; use two colours instead of three (\textit{i.e.\/} reduce the gauge group from SU(3) to SU(2)); discretize space and time (go to the lattice formulation); and eventually investigate the problem in lower-dimensional spacetime.

	Cut to the bone, in (3+1)-dimensional SU(2) Yang--Mills theory the problem is to find the lowest-energy eigenstate of the temporal-gauge hamiltonian
satisfying
\onelineequation{SchR}{\int d^3x\left(-\oh\frac{\delta^2}{\delta A^a_k(x)^2}+\of F^a_{ij}(x)^2\right)\Psi_0[A]=E_0\Psi_0[A]}%
together with the Gau\ss-law constraint
\onelineequation{Gauss}{\left(\delta^{ac}\partial_k+g\varepsilon^{abc} A^b_k(x)\right)\frac{\delta\Psi_0[A]}{\delta A^c_k(x)}=0.}%
Albeit simply looking, attempts to solve the equation can claim at most only partial successes. There are, however, a few things that are known about the solution already for decades:

	1.~If we set $g\to 0$, the Schr\"odinger equation reduces to that of (3 copies of) electrodynamics and the solution is well-known:
\onelineequation{g0}{\Psi_0[A]\ \ 
{\stackrel{{g=0}}{=}}\ \ 
{\cal{N}}\exp\left[-\frac{1}{4}{\displaystyle\int} d^3x\;d^3y\; 
F^a_{ij}(x)\;{\displaystyle
\left(\frac{\delta^{ab}}{\sqrt{-\nabla^2}}\right)_{xy}}F^b_{ij}(y)\right].}

	2.~The ground state must be gauge-invariant. The simplest form one can imagine that reduces to Eq.~(\ref{g0}) in the free-field limit is
\onelineequation{GIVWF}{\Psi_0[A]\;
=\;
{\cal{N}}\exp\left[-\frac{1}{4}{\displaystyle\int} d^dx\;d^dy\; 
F^a_{ij}(x)\;{\displaystyle
{{\cal{K}}^{ab}_{xy}}[-{\cal{D}}^2]}\;F^b_{ij}(y)\right]}
with some kernel ${\cal{K}}$ depending on ${\cal{D}}^2$ (the covariant laplacian in the colour adjoint representation), and fulfilling
\onelineequation{limitK}{\lim_{g\to0}{\cal{K}}^{ab}_{xy}[-{\cal{D}}^2]=\left(\frac{\delta^{ab}}{\sqrt{-\nabla^2}}\right)_{xy}.}
In fact, all proposals of the VWF that will be confronted with numerical data in this paper are of the above form.

	3.~It was suggested \cite{Greensite:1979yn,Halpern:1978ik,Kawamura:1996us} that for sufficiently long-wavelength, slowly varying gauge fields the VWF has the following, so called \textit{dimensional-reduction} form:
\onelineequation{DR}{\Psi_0[A]=\cN\exp\left(-\oh\mu\int d^3x\;\mbox{Tr}[F^2_{ij}(x)]\right)\quad\dots\quad\underline{\mbox{DR}}}%
 This form, \textit{a.k.a.\/}\ the magnetically disordered vacuum, leads incorrectly \textit{e.g.\/}\ to exact Casimir scaling of potentials between coloured sources, so it cannot be valid for \textit{arbitrary} gauge fields.

	The problem of finding the Yang--Mills VWF has been addressed by various techniques.
\footnote{See \textit{e.g.\/}\ Sec.\ II of Ref.~\cite{Greensite:2011pj} and references therein, for the most recent work consult  Ref.~\cite{Krug:2013yq}.} Some proposals for the VWF will be reviewed in Sec.~\ref{section2}. Then I will present (Sec.~\ref{section3}) a method for computing relative weights of various gauge-field configurations in numerical simulations of the Yang--Mills theory in the lattice formulation. Some results will be presented in Sec.~\ref{section4}.  Sec.~\ref{section5} summarizes pluses and minuses of the present approach.

\section{As You Like It (or As We Like It)}\label{section2}
\sectionquote{which introduces some popular Ans\"atze and provides
some justification for one that we like most.}

	Head-on attempts to solve Eqs.~(\ref{SchR}) and (\ref{Gauss}), \textit{e.g.\/}\ by weak-coupling expansion in powers of~$g$, quickly run into complicated  intractable expressions (see \cite{Krug:2013yq}). Some approaches tried instead to bridge the gap between the free-field limit (\ref{g0}) and the dimensional-reduction form of Eq.~(\ref{DR}) by educated guesses of the interpolating approximate vacuum wave functional. 

	Almost 20 years ago, Samuel~\cite{Samuel:1996bt} proposed a simple expression of the type (\ref{GIVWF})
\onelineequation{Samuel}{\Psi_0[A]\;
{\stackrel{}{=}}\;
{\cal{N}}\exp\left[-\of{\displaystyle\int} d^3x\;d^3y\; 
F^a_{ij}(x)\;{\displaystyle
\left(\frac{1}{\sqrt{-{\cal{D}}^2+m^2_0}}\right)^{ab}_{xy}}F^b_{ij}(y)\right]}
and estimated with its use the $0^{++}$ glueball mass. However, there may be a problem with this Ansatz: the operator $(-{\cal{D}}^2)$ has a positive definite spectrum, finite with a lattice regularization, and lattice simulations indicate that its lowest eigenvalue $\lambda_0$ tends to infinity for typical configurations in the continuum limit. This is illustrated in Fig.~\ref{l0color}. 
\onefigure{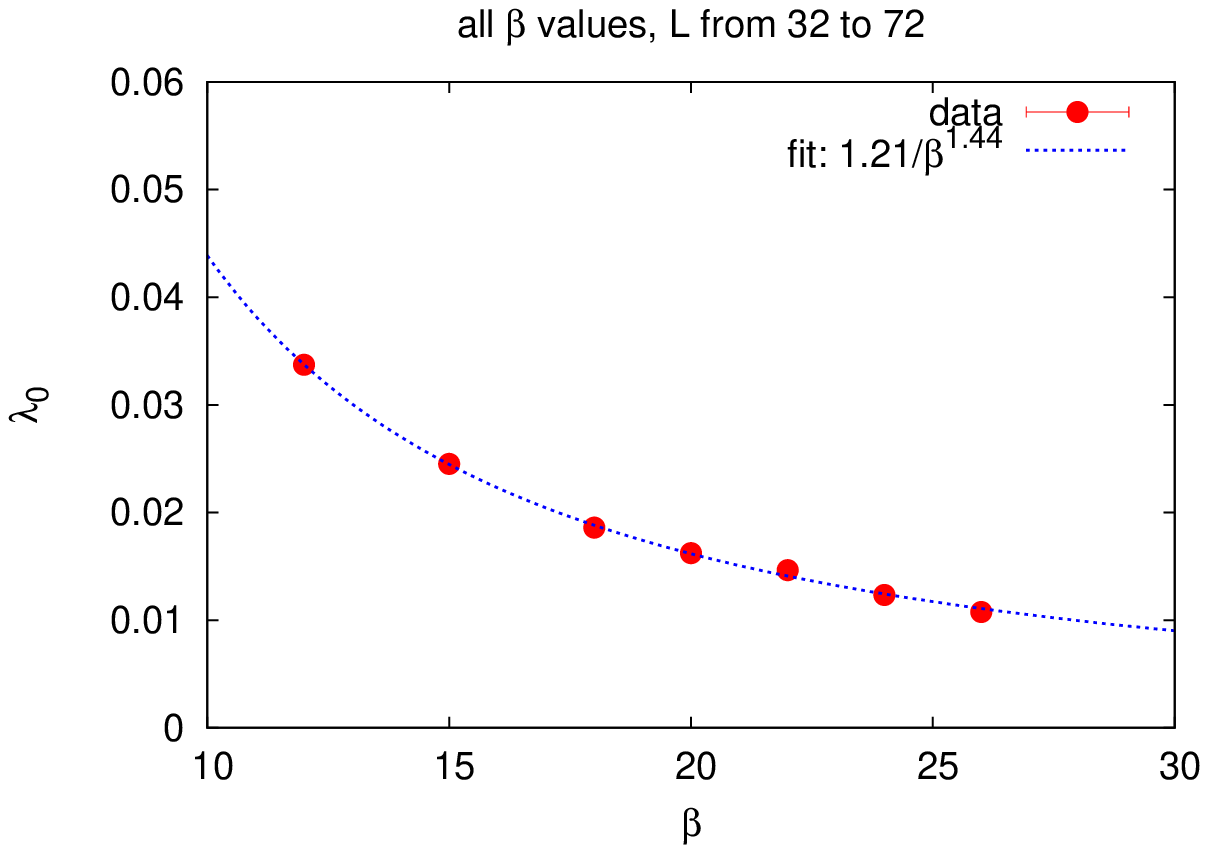}{0.5\textwidth}{$\lambda_0$ vs.\ $\beta$ from simulations of SU(2) lattice gauge theory in $(2+1)$ dimensions at various couplings and lattice volumes. The best fit to data is $\lambda_0\propto \beta^{-1.4}$, which differs from the expected $\beta^{-2}$ dependence and indicates that $\lambda_0$ diverges in the continuum limit.}{t!}

	We therefore proposed to subtract from $(-{\cal{D}}^2)$ its lowest eigenvalue, resulting in the approximate VWF \cite{Greensite:2007ij}:
\onelineequation{GO}{\Psi_0[A]=\cN\exp\left[-\of\int d^3x\;d^3y\;F^a_{ij}(x)\left(\frac{1}{\sqrt{-{\cal{D}}^2[A]-\lambda_0+m^2}}\right)^{ab}_{xy} F^b_{ij}(y) \right]\quad\dots\quad\underline{\mbox{GO}}}%
with $m$ being a free (mass) parameter. This expression is assumed to be regularized by a lattice cut-off, and we use the simplest discretized form of $(-{\cal{D}}^2)$:
\onelineequation%
{}{\left({-{\cal D}^2}\right)^{ab}_{xy}=
\displaystyle\sum_{k=1}^3 \left[2\delta^{ab}\delta_{xy}-
{\cal U}^{ab}_k(x)\delta_{y,x+\hat{k}}-{\cal U}^{\dagger ba}_k(x-\hat{k})\delta_{y,x-\hat{k}}\right],}
where ${{\cal U}^{ab}_k(x)=\frac{1}{2}\mbox{Tr}\left[\sigma^a U_k(x) \sigma^b U^\dagger_k(x)\right]}$, and $U_k(x)$ are the usual link matrices in the fundamental representation.

	An expression analogous to Eq.~(\ref{GO})  in $(2+1)$ dimensions was demonstrated to be a fairly good approximation to the true ground state of the theory by:

-- analytic arguments \cite{Greensite:2007ij},

-- direct computation of some physical quantities in ensembles of true Monte Carlo configurations and those distributed according to the square of the GO VWF \cite{Greensite:2007ij,Greensite:2010tm}, and

-- consistency of measured probabilities of test configurations  with expectations based on the proposed VWF~\cite{Greensite:2011pj}.

	The most sophisticated attempt to compute the VWF analytically in $(2+1)$ dimensions was undertaken by Karabali, Kim, and Nair~\cite{Karabali:1998yq}. They reformulated the theory with help of new gauge-invariant variables, and solved the Yang--Mills Schr\"odinger equation approximately for the VWF in their terms. They argue that, when expressed back in the old variables, this VWF assumes the form:
\onelineequation{KKNngi}{\Psi_0[A]=\cN\exp\left[-\oh\int d^2x\;d^2y\;B^a(x)\left(\frac{1}{\sqrt{-\nabla^2+m^2}+m}\right)_{xy} B^b(y) \right],}
This is by itself \textit{not} gauge-invariant, but can be made such along the lines of Eqs.~(\ref{GIVWF}) and (\ref{GO}) by replacing the ordinary laplacian by the covariant laplacian in the adjoint representation, with a~$\lambda_0$~subtraction:
\onelineequation{KKN3}{\Psi_0[A]=\cN\exp\left[-\of\int d^3x\;d^3y\;F^a_{ij}(x)\left(\frac{1}{\sqrt{-{\cal{D}}^2[A]-\lambda_0+m^2}+m}\right)^{ab}_{xy} F^b_{ij}(y) \right]\quad\dots\quad\underline{\mbox{KKN}}}
Such an expression, however, has never been proposed by the authors of Ref.~\cite{Karabali:1998yq} in their papers, and represents only yet another interpolating VWF of the type~(\ref{GIVWF}) that can be confronted with our numerical data. 

\section{Measure for Measure}\label{section3}
\sectionquote{wherein is shown how one can measure \textbf{\textit{``nothing''}} and
learn from it \textbf{\textit{something}}.}

	The squared VWF could, at least in principle, be computed on a lattice by evaluating the path integral (written below only symbolically, with $\delta_\mathrm{t.g.f.}$ imposing the temporal gauge):
\onelineequation{PI}{\Psi^2_0[U']=\frac{1}{Z}\int [DU]\;\delta_\mathrm{t.g.f.}\;\prod_{\mathbf{x},i}\delta[U_i(\mathbf{x},0)-U'(\mathbf{x})]e^{-S[U]}.}%
\noindent
An integral of this type is, however, difficult to estimate numerically, because of the $\delta$-functions. The method that enables one to compute -- simply and directly -- ratios  $\Psi^2[U^{(n)}]/\Psi^2[U^{(m)}]$  for some test configurations was proposed by Greensite and Iwasaki \cite{Greensite:1989aa}. Their \textit{{relative-weight method}} consists of the following: Take a finite set of gauge-field configurations ${\mathcal{U}}=\lbrace U_i^{(j)}(\mathbf{x}),j=1,2,\dots,M\rbrace$ (assuming they lie near to each other in the configuration space). One puts \textit{e.g.\/}\ the $j=1$ configuration on the $t=0$ plane, and runs Monte Carlo simulations with the usual update algorithm (\textit{e.g.\/}\ heat-bath) for all spacelike links at $t\ne0$ and for timelike links. The spacelike links at $t=0$ are, after a certain number of sweeps, updated all at once selecting one configuration from the set $\mathcal{U}$ at random and accepting/rejecting it via the Metropolis algorithm. Then\
\onelineequation{ratio}{\frac{\Psi^2[U^{(n)}]}{\Psi^2[U^{(m)}]}=\lim_{N_\mathrm{tot}\to\infty}\frac{N_n}{N_m}
=\lim_{N_\mathrm{tot}\to\infty}\frac{N_n/N_\mathrm{tot}}{N_m/N_\mathrm{tot}},}%
where $N_n$ ($N_m$) is the number of times the $n$-th ($m$-th) configuration is accepted and $N_\mathrm{tot}$ is the total number of updates.

	The VWF can always be written in the form 
\onelineequation{}{\Psi^2[U]={\mathcal{N}}e^{-R[U]}.}%
According to Eq.~(\ref{ratio}), the measured values of $-\log(N_n/N_\mathrm{tot})$ should fall on a straight line with unit slope as functions of $R[U^{(n)}]$, see Fig.~\ref{prob_k_2_2_l20_c_prob_k_2_5_l20_c} for examples.
\twofigures{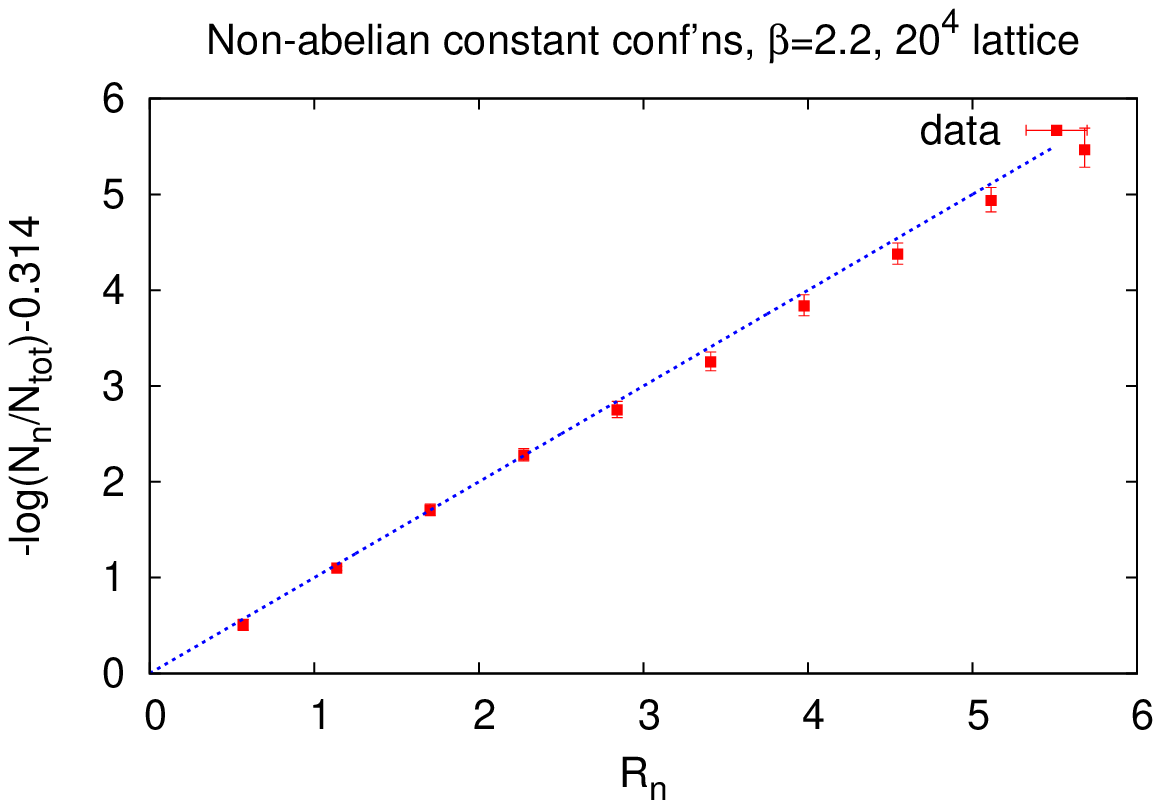}{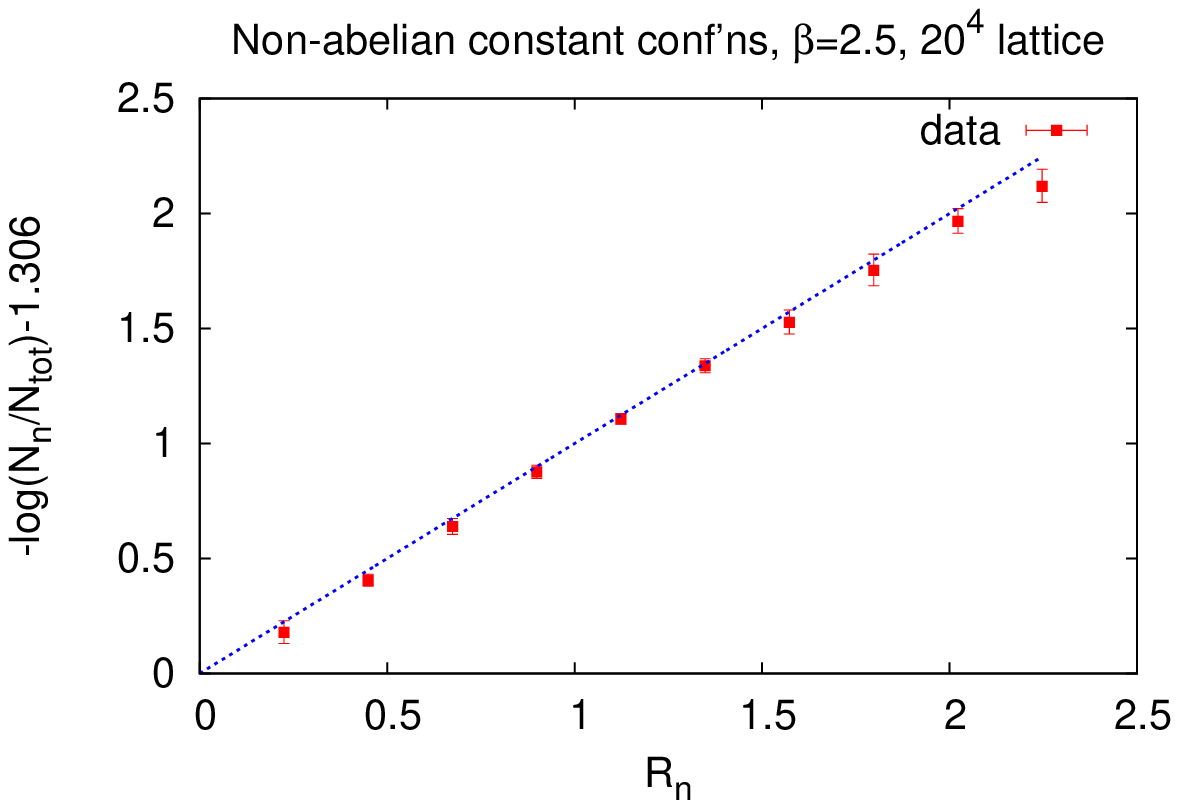}{0.48\textwidth}{$-\log(N_n/N_\mathrm{tot})$ (shifted by constant) vs.\ $R_n=\mu\kappa n$ for ${\mathcal{U}}_\mathrm{NAC}$ [\textit{cf.\/}\ Eq.~(\protect\ref{NAC}) below] with $\kappa=0.14$, on $20^4$ lattice. The values of $\mu$ come out to be $4.06 (4)$ and $1.60 (2)$ for $\beta=2.2$ and $2.5$, respectively.}{b!}

	We have performed numerical simulations using the relative-weight method for two kinds of simple gauge-field configurations.

\medskip
	\textbf{\textit{1.~Non-abelian constant configurations\/}:}
\onelineequation{NAC}{{\cal U}_\mathrm{NAC}=\left\{U_k^{(n)}(x)=\sqrt{1-\left(a^{(n)}\right)^2}\mathbf{1}+ia^{(n)}\bm{\sigma}_k\right\},} 
where
\onelineequation{NAC_a}{a^{(n)}=\left(\frac{\kappa}{6L^3}n\right)^{1/4},\qquad n=1,2,\dots, 10.}%
For NAC configurations one expects:
\onelineequation{NACfit}{-\log(N^{(n)}/N_\mathrm{tot})=R^{(n)}+ \mbox{const.}=\kappa n\times{\mu}+ \mbox{const.}}
The constant $\kappa$, regulating amplitudes of these configurations, is chosen so that the ratio $R^{(10)}/R^{(1)}$ is not too small, ${\cal{O}}(10^{-4}\div10^{-3})$, otherwise the Metropolis updates would hardly accept configurations with higher $n$. 

\medskip
	\textbf{\textit{2.\ Abelian plane-wave configurations}:} 
\onelineequation{APW}{{\cal U}_\mathrm{APW}=\left\{U_1^{(j)}(x)=\sqrt{1-\left(a^{(j)}_\textbf{\textit{n}}(x)\right)^2}\mathbf{1}+ia^{(j)}_\textbf{\textit{n}}(x)\bm{\sigma}_3,\quad U_2^{(j)}(x)=U_3^{(j)}(x)=\mathbf{1}\right\}, }
where $\textbf{\textit{n}}=(n_1,n_2,n_3)$, and
\onelineequation{APW_a}{a^{(j)}_\textbf{\textit{n}}=\sqrt{\frac{\alpha_\textbf{\textit{n}}+\gamma_\textbf{\textit{n}}j}{L^3}}\cos\left(\frac{2\pi}{L}\textbf{\textit{n}}\cdot\textbf{\textit{x}}\right),\qquad j=1,2,\dots, 10.}
	Again, pairs of $(\alpha_\textbf{\textit{n}},\gamma_\textbf{\textit{n}})$ characterizing abelian plane waves with the wavenumber $\textbf{\textit{n}}$ in the above equations were carefully selected so that the actions of plane waves with different $j$ were not much different (to ensure reasonable Metropolis acceptance rates in the method described above).

	The expectation for APW configurations is
\onelineequation{APWfit}{\displaystyle-\log(N^{(j)}_\textbf{\textit{n}}/N_\mathrm{tot})=R^{(j)}_\textbf{\textit{n}}+ \mbox{const.}=
\oh(\alpha_\textbf{\textit{n}}+\gamma_\textbf{\textit{n}}j)\times{\omega(\textbf{\textit{n}})}+ \mbox{const.}}

\section{The Comedy of Errors}\label{section4}
\sectionquote{which showcases some results, discusses pitholes, and
compares the results to the Ans\"atze.}

	Our aim is to compare computed relative weights of non-abelian constant and abelian plane-wave configurations with predictions of the DR, GO, and KKN-inspired wave functionals discussed in Section \ref{section2}. NAC configurations are not useful for that purpose. However, they served for ``calibrating'' our computer code by comparison with the results of Ref.~\cite{Greensite:1989aa}, obtained on lattices of much smaller size. For a number of $\beta$ values we determined the slope $\mu$ in Eq.~(\ref{NACfit}). Our data from $16^4$ and $20^4$ lattices clearly agree with those of Ref.~\cite{Greensite:1989aa} from $6^4$ and $8^4$. At small $\beta$ the strong-coupling prediction $\mu(\beta)=\beta$ is confirmed, in the scaling window $\mu(\beta)$ behaves as a physical quantity with the dimension of inverse mass:
\onelineequation{mu_phys}{\mu(\beta)f(\beta)=\mu_\mathrm{phys}\approx 0.0269(3),}
where
\onelineequation{fbeta}{f(\beta)=\left(\frac{6\pi^2\beta}{11}\right)^\frac{51}{121}\exp\left(-\frac{3\pi^2\beta}{11}\right).} 
\onefigure{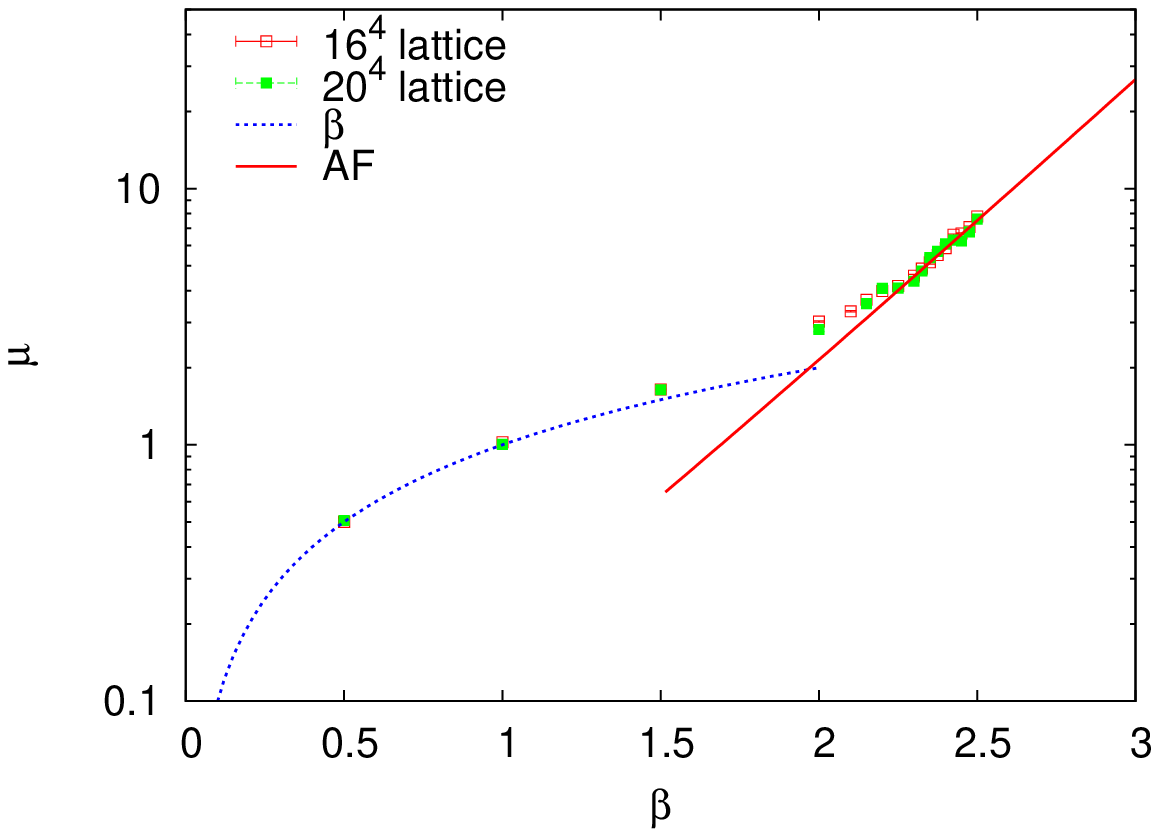}{0.5\textwidth}{Variation of $\mu$ with $\beta$, estimated from data for NAC configurations on $16^4$ and $20^4$ lattices.}{t!}

	For a particular set of abelian plane waves with the wavenumber $\textbf{\textit{n}}$  one can determine the slope $\omega(\textbf{\textit{n}})$ from the measured values of relative weights of individual plane waves by a fit of the form~(\ref{APWfit}). The expected linear dependence was observed with all our data at all couplings, wave numbers, and parameter choices; for examples see Fig.~\ref{prob_k_a1_2p4_k010_l24_c_prob_k_a1_2p5_k015_l30_c}.
\twofigures{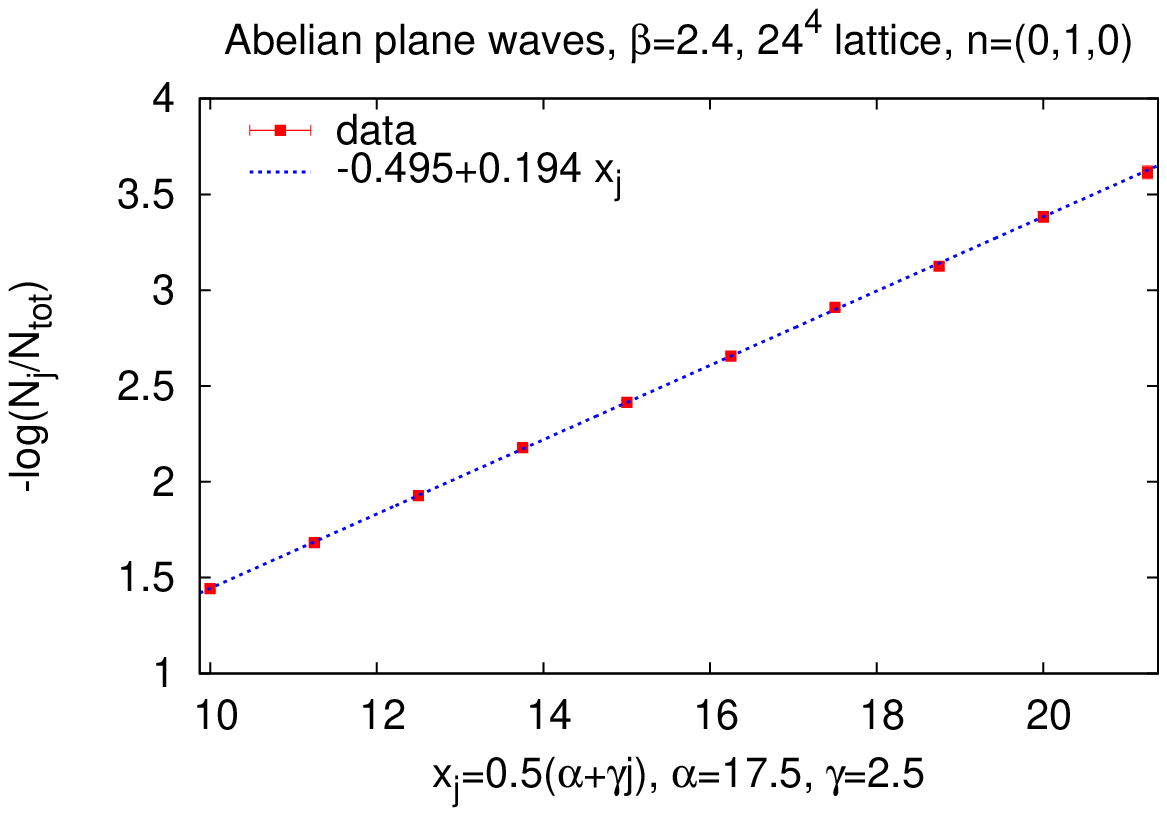}{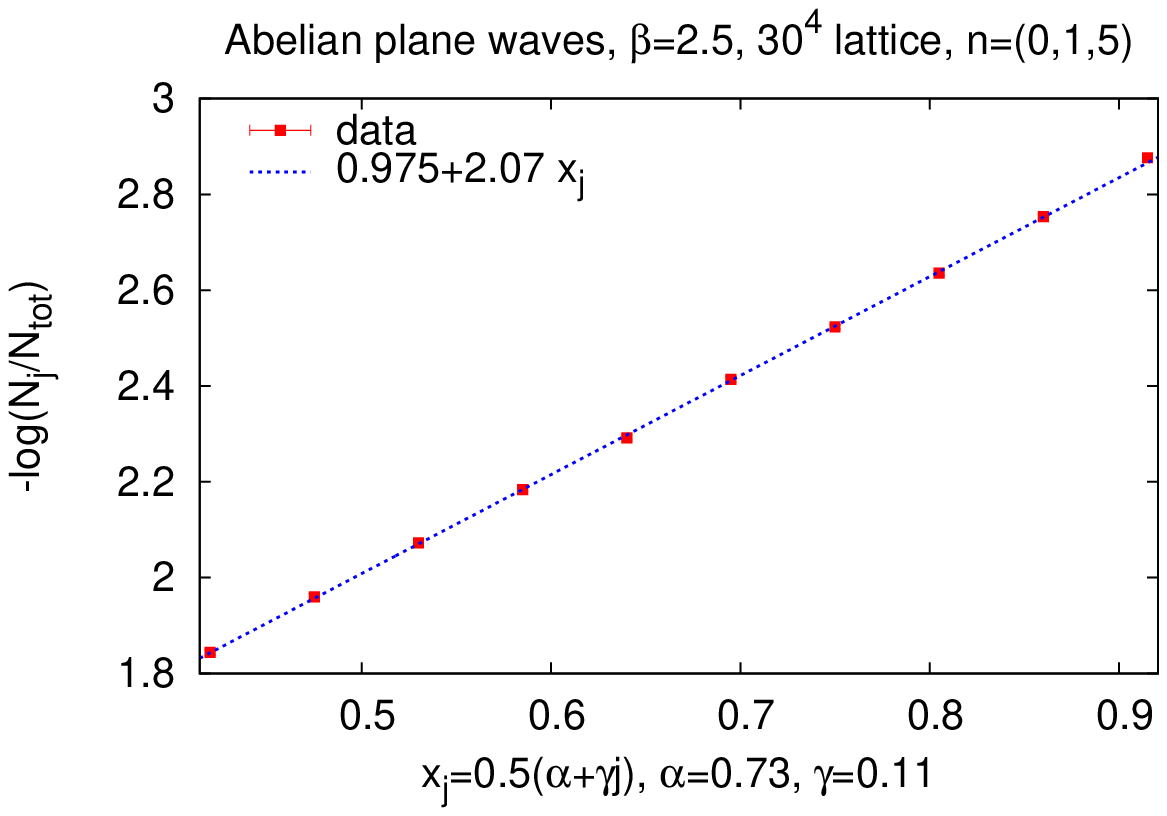}{0.48\textwidth}{$-\log(N^{(j)}_\textbf{\textit{n}}/N_\mathrm{tot})$ vs.\ $\oh(\alpha_\textbf{\textit{n}}+\gamma_\textbf{\textit{n}}j)$ for ${\mathcal{U}}_\mathrm{APW}$ [see Eq.~(\protect\ref{APW})]. }{b!}
However, one could imagine that the dependence is linear only \textit{locally}, in a certain narrow window, and the slope $\omega(\textbf{\textit{n}})$ could depend strongly on the choice of parameters $(\alpha_\textbf{\textit{n}},\gamma_\textbf{\textit{n}})$. This does not seem to be the case, as exemplified in Fig.~\ref{prob_for_more_data_sets}.
\onefigure{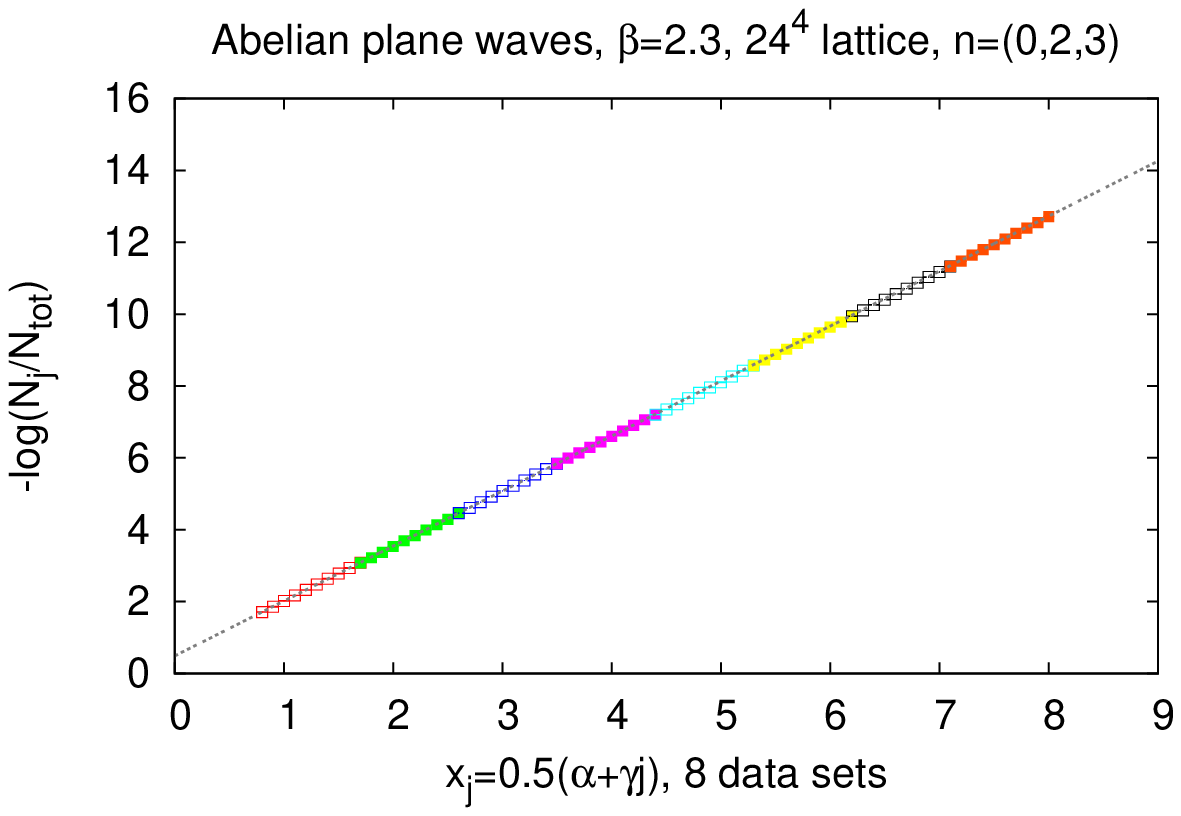}{0.5\textwidth}{The slope determined from $-\log(N_j/N_\mathrm{tot})$ does not strongly depend on the choice of parameters $\alpha$ and $\gamma$ for abelian plane waves.  Eight sets of configurations at a given $\beta$ and wave-number are superimposed here; the last configuration in one set had identical amplitude with the first configuration in the next set. The measured mean values of $-\log(N_j/N_\mathrm{tot})$ were renormalized so that the value for the last configuration in one set coincided with that of the first configuration in the next set. The straight-line fit shown in the figure comes from the first data set (red open squares). The slopes obtained from other sets almost do not differ, the variation is at most 1\%, between 1.523 and 1.539.}{t!}

	The dependence of $\omega(\textbf{\textit{n}})$  on $\textbf{\textit{n}}$ can now be compared with expectations based on the DR, GO, and KKN-inspired VWFs. We performed the following fits:
\onelineequation{fits}{{\omega(\textbf{\textit{n}})}=\left\{
\begin{array}{l c l}
a+{b}k^2(\textit{\textbf{n}}) & \qquad\dots\qquad & \underline{\mbox{DR}},\\[2mm]
{\displaystyle{c}\frac{k^2(\textit{\textbf{n}})}{\sqrt{k^2(\textit{\textbf{n}})+{m}^2}}} & \dots & \underline{\mbox{GO}},\\[2mm]
{\displaystyle{c}\frac{k^2(\textit{\textbf{n}})}{\sqrt{k^2(\textit{\textbf{n}})+{m_1}^2}+{m_2}}} & \dots & \underline{\mbox{inspired by KKN}},
\end{array}\right.}%
where
\onelineequation{momentum}{k^2(\textit{\textbf{n}})=2\sum_i\left(1-\cos\frac{2\pi n_i}{L}\right).}
In the KKN-inspired fit we introduced two fit mass parameters, $m_1$ and $m_2$, instead of just $m$, \textit{cf.}\ Eq.~(\ref{KKN3}). We then performed a fit with both parameters free, and a constrained fit with $m_1=m_2$. It turned out that the former had a lower $\chi^2$ and the preferred value of $m_1$ was close to 0.

	Prototype plots for fits of the form (\ref{fits}) are displayed in Fig.~\ref{omega_2_5_l30_c_omega_2_5_l30_KKN_c}  for the DR and GO forms (left panel), and for the KKN-inspired forms (right panel). All forms in Eq.~(\ref{fits}) describe the data reasonably at low plane-wave momenta, none of them is satisfactory for larger momenta.
\twofigures{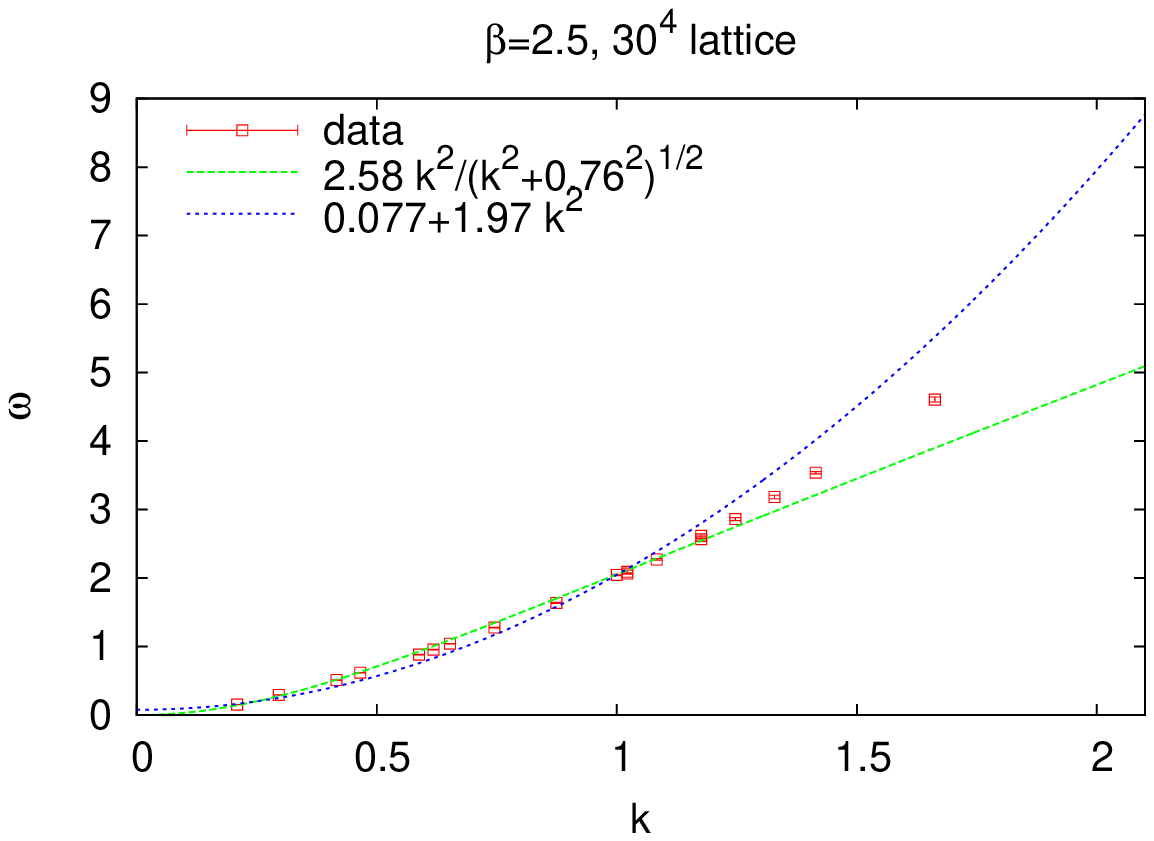}{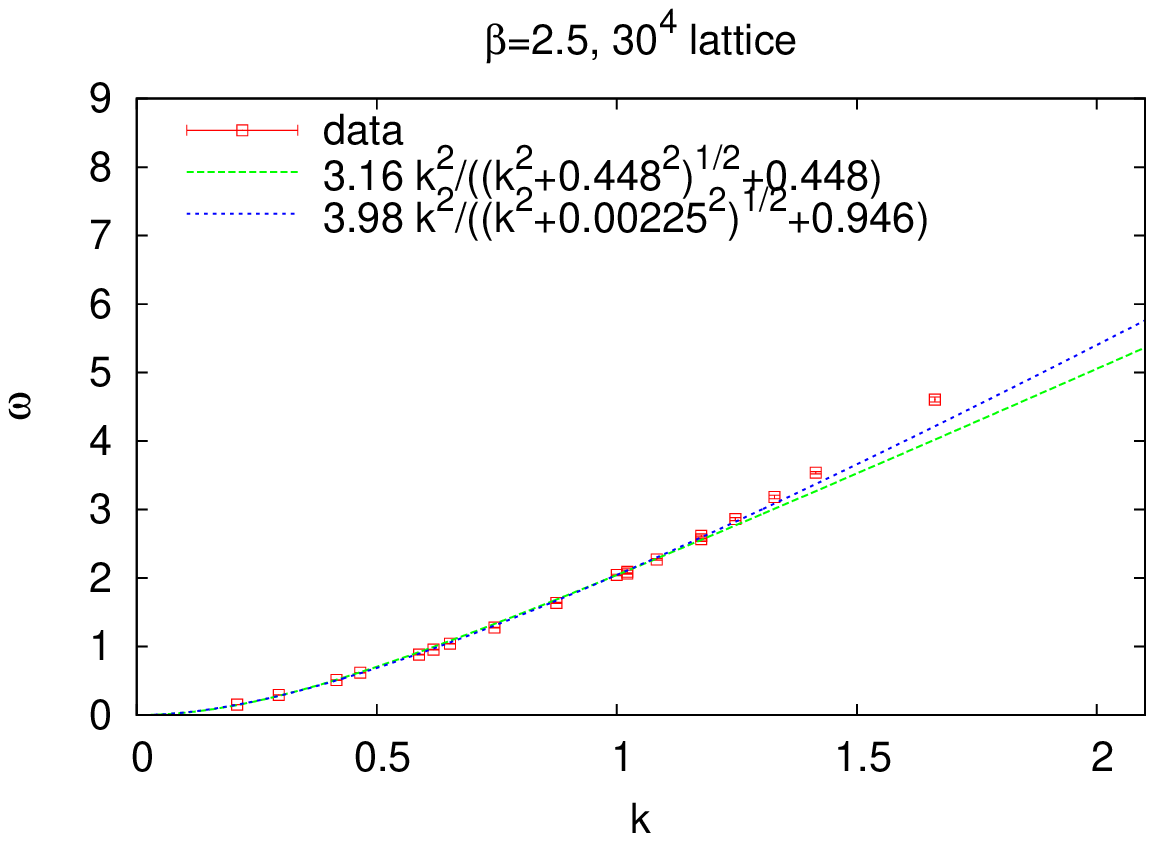}{0.48\textwidth}{$\omega(\textbf{\textit{n}})$ vs.\ $k(\textbf{\textit{n}})$ for ${\cal U}_\mathrm{APW}$ sets, with the DR and GO fits (left), and ``KKN-inspired'' fits (right).}{t!}

	The agreement with data greatly improves at all couplings by adding another parameter $d$ to the GO form:
\onelineequation{best}{{\omega(\textbf{\textit{n}})}={c}\frac{k^2(\textit{\textbf{n}})}{\sqrt{k^2(\textit{\textbf{n}})+{m}^2}}\left[1+{d}k(\textit{\textbf{n}})\right].}%
see Fig.~\ref{omega_2_5_l30_guess_c}. This would correspond in the continuum limit to the following choice of the kernel in~(\ref{GIVWF}):
\onelineequation{guess-kernel}{{\cal{K}}^{ab}_{xy}[-{\cal{D}}^2]\propto\left(\frac{1}{\sqrt{-{\cal{D}}^2-\lambda_0+m_\mathrm{phys}^2}}+{d_\mathrm{phys}}
\sqrt{\frac{-{\cal{D}}^2-\lambda_0}{{-{\cal{D}}^2-\lambda_0+m_\mathrm{phys}^2}}}\right)^{ab}_{xy}.}
\onefigure{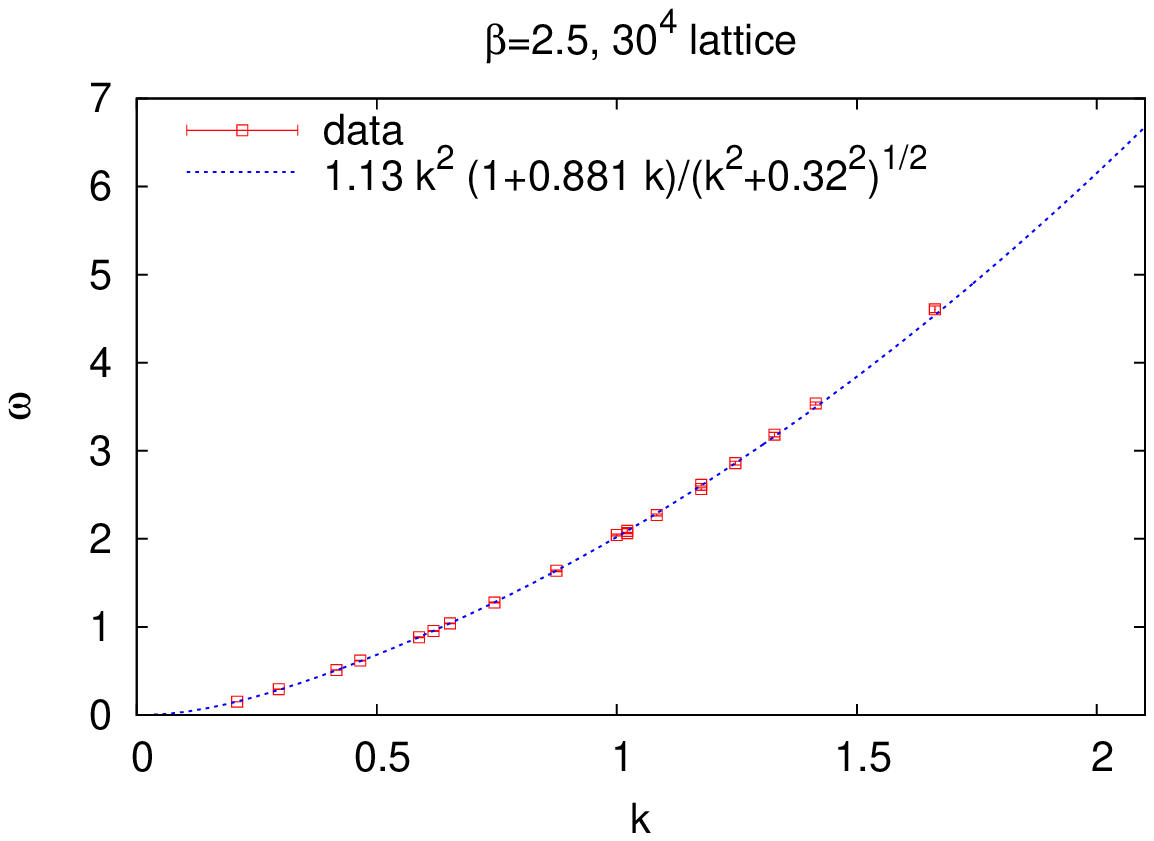}{0.5\textwidth}{$\omega(\textbf{\textit{n}})$ vs.\ $k(\textbf{\textit{n}})$ for ${\cal U}_\mathrm{APW}$ sets, with the best fit of the form (\protect\ref{best}).}{b!} 

\begin{figure}[p]
\centering
\includegraphics[width=0.5\textwidth]{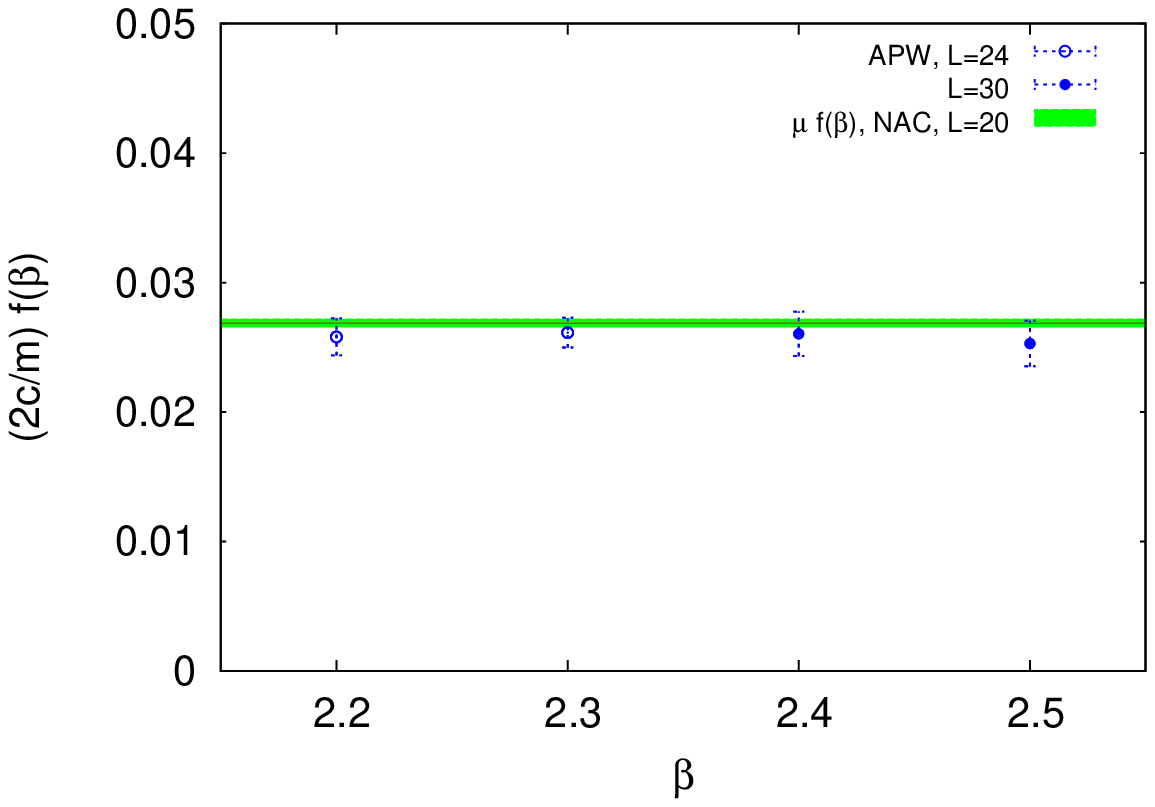}
\caption{The combination $(2c/m)f(\beta)$ of the best fit to data, Eq.~(\protect\ref{best}). Also displayed is $\mu f(\beta)=0.0269(3)$ derived from non-abelian constant configurations.}\label{scaling1}
\end{figure}%

\begin{figure}[p]
\centering
\begin{tabular}{c c}
\includegraphics[width=0.48\textwidth]{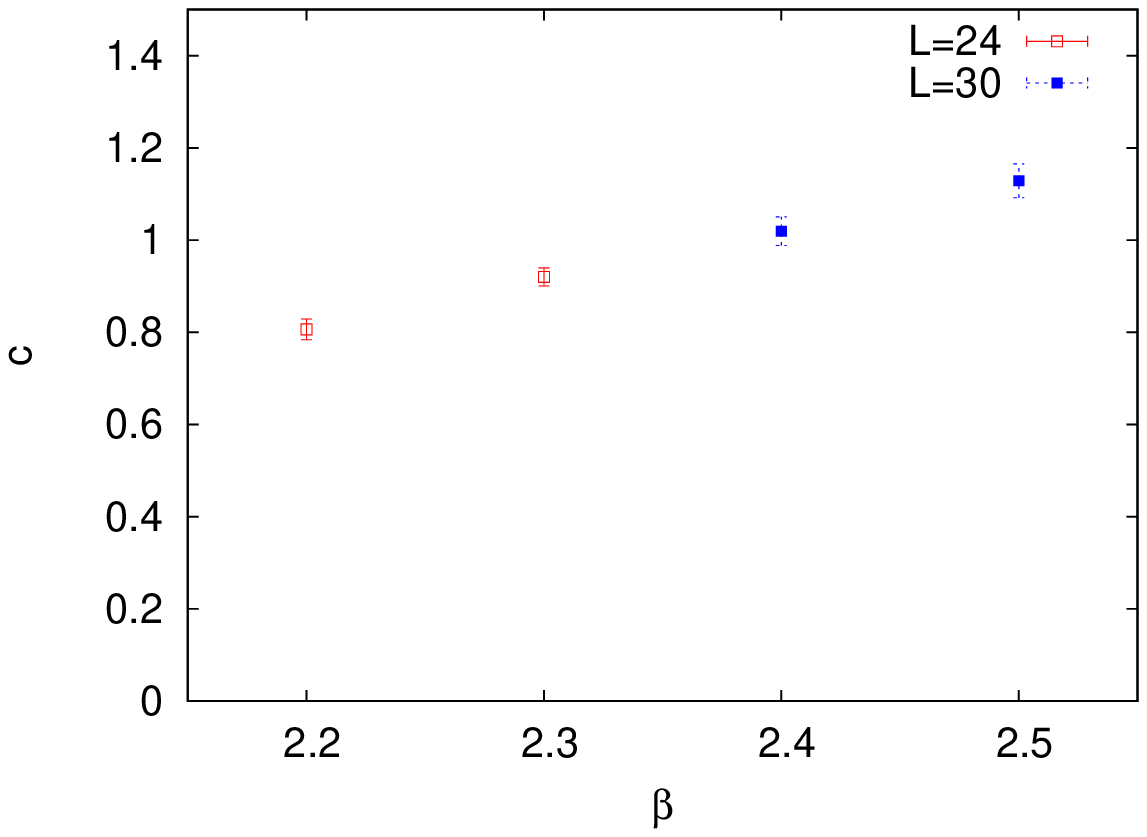}&\includegraphics[width=0.48\textwidth]{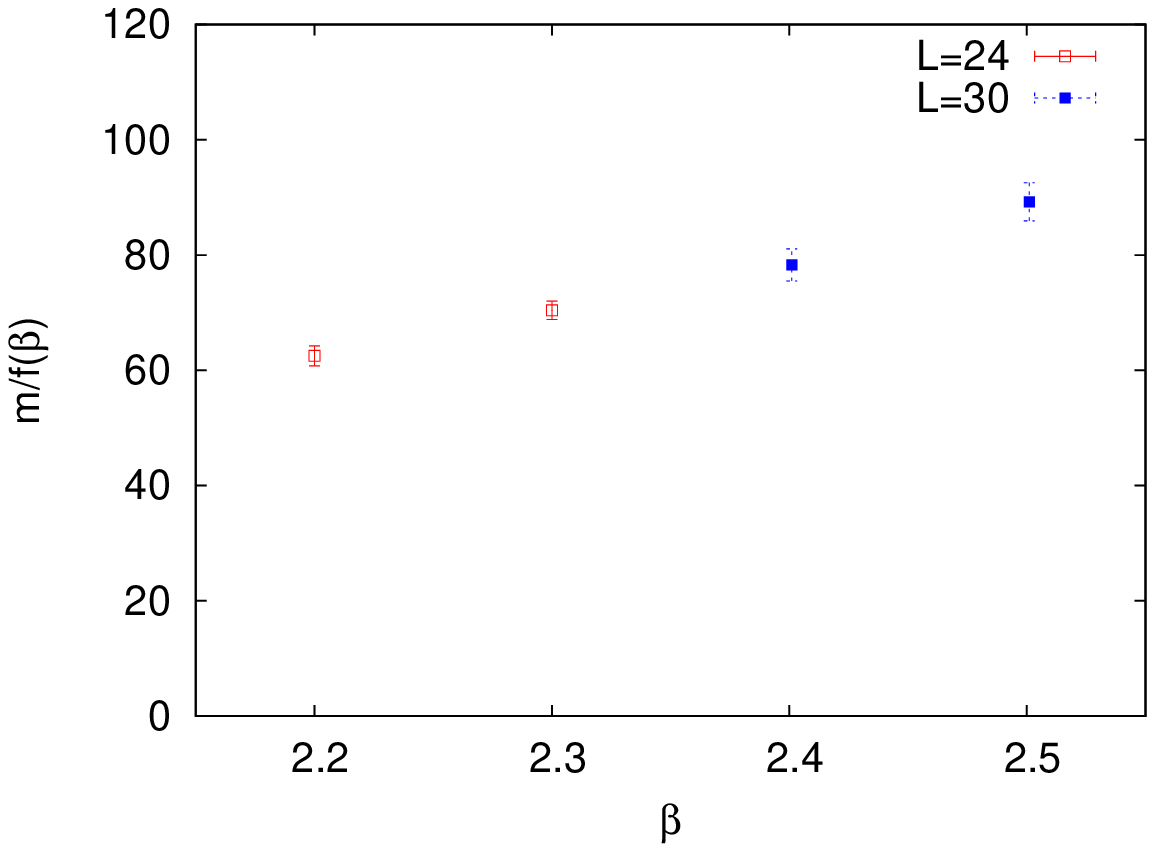}
\end{tabular}
\caption{The parameter $c$ (left), and the rescaled parameter $m/f(\beta)$ (right) of the best fit, Eq.~(\protect\ref{best}), vs.~$\beta$.}\label{scaling2}
\end{figure}%

\begin{figure}[p]
\centering
\includegraphics[width=0.5\textwidth]{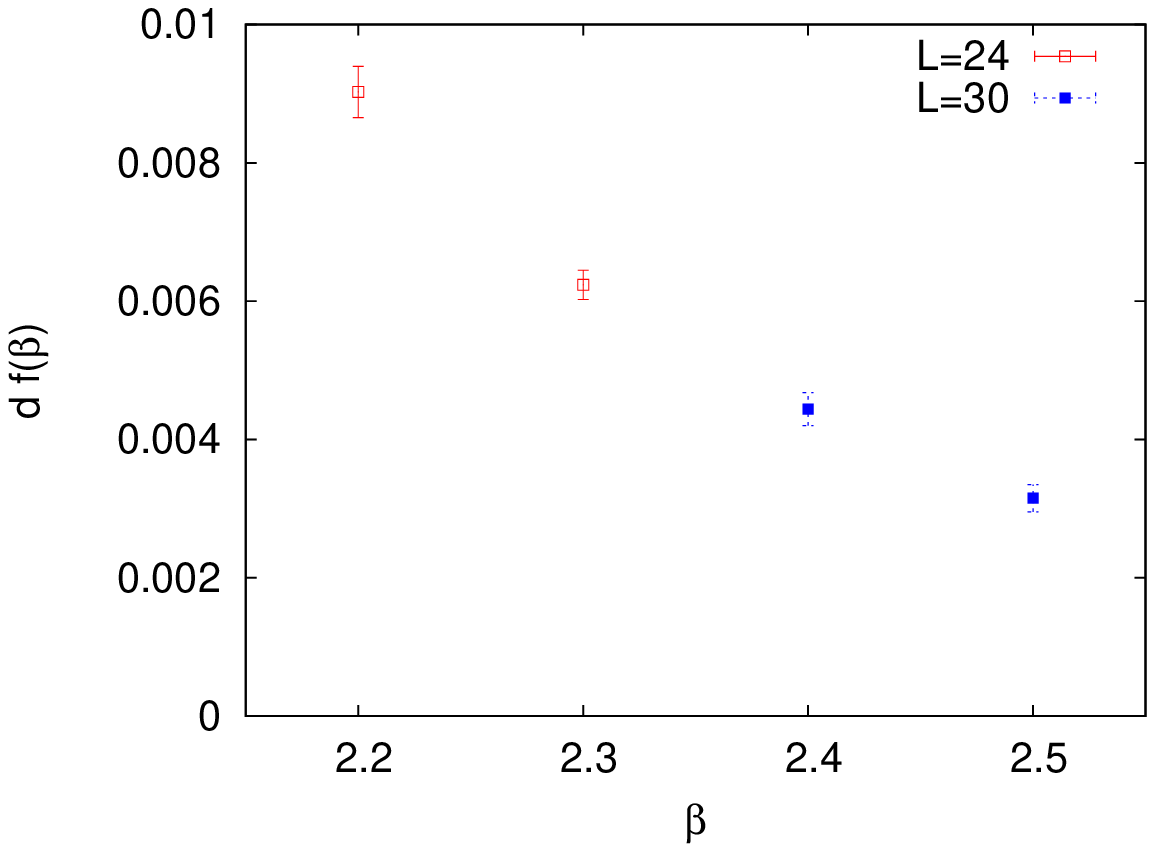}
\caption{The rescaled parameter $d f(\beta)$ of the best fit, Eq.~(\protect\ref{best}), vs.~$\beta$.}\label{scaling3}
\end{figure}%

	For small-amplitude constant configurations the forms of the VWF in Eqs.~(\ref{DR}) and (\ref{GO}) coincide. It is therefore an important consistency check whether the value of $\mu_\mathrm{NAC}$ determined from sets of non-abelian constant configurations agrees with the appropriate combination of parameters obtained for abelian plane waves. In particular, one expects:
\onelineequation{consistency}{\mu_\mathrm{NAC}=\left(\frac{2c}{m}\right)_\mathrm{APW}.}
As seen convincingly in Fig.~\ref{scaling1}, our results clearly pass this nontrivial check.

	If the parameters of the best fit, Eq.~(\ref{best}), correspond to physical quantities in the continuum limit, they should scale correctly when multiplied by the appropriate power of the function $f(\beta)$, Eq.~(\ref{fbeta}). The behaviour of $[2c(\beta)/m(\beta)]f(\beta)$, $c(\beta)$, $m(\beta)/f(\beta)$, and $d(\beta)f(\beta)$ vs.\ the coupling~$\beta$ is displayed in Figs.~\ref{scaling1}, \ref{scaling2} and \ref{scaling3}. While the scaling of $(2c/m)$ is almost perfect (Fig.~\ref{scaling1}),  it is not convincing for $c$ and $m$ separately (Fig.~\ref{scaling2}), though the variation over the range of $\beta = 2.2\div2.5$ is not so large. On the contrary, $d(\beta)f(\beta)$ falls down considerably over the same range (Fig.~\ref{scaling3}). The data thus indicate that the physical value of $d$ vanishes in the continuum limit. This suggests an idea that the form of the VWF, Eq.~(\ref{GO}), proposed in Ref.~\cite{Greensite:2007ij}, might be recovered in the continuum limit.

\section{All's Well That Ends Well (?)}\label{section5}
\sectionquote{wherein some optimistic and pessimistic conclusions are
formulated.}

	Let's group the messages of this work into two categories:
\begin{center}
\begin{tabular}{p{0.46\textwidth} p{0.0\textwidth} p{0.46\textwidth}}
\centerline{\fbox{\textbf{\textit{Pluses}}}}&&\centerline{\fbox{\textbf{\textit{Minuses}}}}\\
There is a method to measure (on a lattice) relative probabilities
of various gauge-field configurations in the Yang--Mills vacuum. && 
The method works reasonably well for configurations rather
close in configuration space.\\[3mm]
Both for nonabelian constant and for long-wavelength abelian plane-wave configurations the measured
probabilities are consistent with the dimensional reduction
form, and the coefficients $\mu$ for these sets agree. &&
Neither the dimensional-reduction form of the vacuum wave functional, nor our proposal, nor the 
forms inspired by the work of Karabali \textit{et al.\/}, describe the data satisfactorily for
larger plane-wave momenta. \\[3mm]
The data are nicely described by a modification of our proposal,
and the correction term may vanish in the continuum
limit.&&
The configurations tested so far, both nonabelian constant and abelian plane-wave configurations, are rather
atypical, not representatives of true vacuum fields.\\[3mm]
&&
One badly needs a method of generating configurations distributed
according to the proposed vacuum wave functionals.\\[3mm]
\end{tabular}
\end{center}

	We presented here only a selection of our results, for more details consult Ref.~\cite{Greensite:2013zz}. Preliminary results were also presented at other conferences \cite{Greensite:2013nb}.

\begin{acknowledgments}
\sectionquote{wherein I thank all who should be thanked, sincerely hoping nobody is forgotten.}

I am grateful to the organizers for arranging this most pleasant and inspiring workshop and for inviting me to participate and present this talk. I acknowledge cooperation with Hugo Reinhardt and Adam Szczepaniak which resulted in Ref.~\cite{Greensite:2011pj}. Pierre van Baal's sentence \textit{``Who thought so much can be said about nothing.''} in the concluding section of his lecture on the QCD vacuum at the Lattice'97 conference \cite{vanBaal:1997vi} inspired the subtitle of my talk. I was lucky that William Shakespeare had written enough comedies to choose my subtitle and section names from. 

This research was supported in part by the U.S.\ Department of Energy under Grant No.\ DE-FG03-92ER40711 (J.G.), by the Slovak Research and Development Agency under Contract No.\ APVV--0050--11, and by the Slovak Grant Agency for Science, Project VEGA No.\ 2/0072/13 (\v{S}.O.). In initial stages of this work,  \v{S}.O.\ was also supported by ERDF OP R\&D, Project meta-QUTE ITMS 2624012002.
\end{acknowledgments}

\end{document}